\documentclass[pdflatex,sn-mathphys-num,iicol]{sn-jnl}

\usepackage{graphicx}%
\usepackage{multirow}%
\usepackage{amsmath,amssymb,amsfonts}%
\usepackage{amsthm}%
\usepackage{mathrsfs}%
\usepackage[title]{appendix}%
\usepackage{xcolor}%
\usepackage{textcomp}%
\usepackage{manyfoot}%
\usepackage{booktabs}%
\usepackage{algorithm}%
\usepackage{algorithmicx}%
\usepackage{algpseudocode}%
\usepackage{listings}%
\usepackage{graphicx}  
\usepackage{dcolumn}   
\usepackage{placeins}
\usepackage{tikz}
\usepackage{lipsum}
\usepackage{soul}
\usepackage{xcolor}
\usepackage{physics}
\usepackage{siunitx}

\usepackage{graphicx}
\usepackage{amsmath}
\usepackage{amsfonts}
\usepackage{epsfig}
\usepackage{slashed}
\usepackage{braket}
\usepackage{comment}
\usepackage{cases}
\usepackage{appendix}
\usepackage{color}
\usepackage{booktabs}
\usepackage{lipsum}
\definecolor{mygreen}{rgb}{0,0.5,0}

\def\bec{\begin{center}}
\def\eec{\end{center}}
\def\beq{\begin{equation}}
\def\eeq{\end{equation}}
\def\bea{\begin{eqnarray}}
\def\eea{\end{eqnarray}}

\usepackage{lineno}

\theoremstyle{thmstyleone}%
\theoremstyle{thmstyletwo}%

\theoremstyle{thmstylethree}%

\begin{document}

\title[Article Title]{Fast Scrambling in the Hyperbolic Ising Model}


\author*[1]{\fnm{Goksu Can} \sur{Toga}}\email{gctoga@ncsu.edu}

\author[2]{\fnm{Abhishek} \sur{Samlodia}}\email{asamlodi@syr.edu}

\author[1]{\fnm{A. F.} \sur{Kemper}}\email{akemper@ncsu.edu}

\affil*[1]{\orgdiv{Department of Physics \& Astronomy}, \orgname{North Carolina State University}, \orgaddress{\city{Raleigh}, \postcode{227695}, \state{NC}, \country{USA}}}

\affil[2]{\orgdiv{Department of Physics}, \orgname{Syracue University}, \orgaddress{\city{Syracuse}, \postcode{13244}, \state{NY}, \country{USA}}}


\abstract{We investigate many-body chaos and scrambling in the Hyperbolic Ising model, a mixed-field Ising model living in the background of $AdS_{2}$. The effect of the curvature is captured by site-dependent couplings obtained from the $AdS_2$ metric applied to a flat nearest-neighbor spin chain. 
Using a combination of out-of-time-ordered correlators (OTOCs), Krylov complexity, and spectral statistics, we present consistent evidence that this model exhibits faster scrambling behavior relative to its flat counterpart. In particular, we observe signatures consistent with fast scrambling dynamics emerging from purely local interactions. 
At the system sizes accessible to tensor network simulations, the OTOCs display short-lived exponential growth regimes, from which we extract effective Lyapunov exponents. These effective finite-size exponents exhibit a temperature dependence broadly compatible with the Maldacena-Shenker-Stanford (MSS) bound within numerical uncertainty. Our results indicate that increasing spatial curvature can significantly decrease scrambling time in systems with only nearest-neighbor interactions, providing a minimal and computationally accessible platform for studying quantum chaos. This makes the model a promising test-bed for exploring scrambling and operator growth in near-term quantum simulation architectures.}





\maketitle
\section{Introduction}\label{sec:intro}
\begin{figure*}
	\centering
	{
		\includegraphics[width=\textwidth]{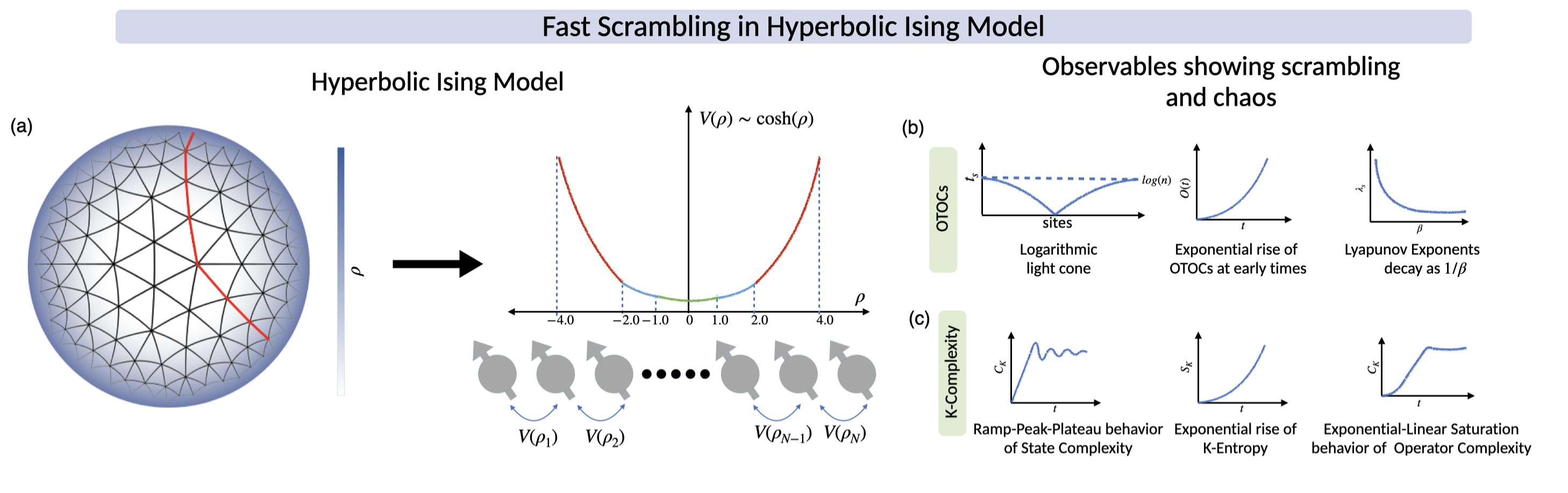}
		\caption{a) Left: Tessellation of the Poincar\'e disk. The red line displays the line we choose to obtain the $AdS_2$ Ising model; the spins are located at the vertices. Right: Site-dependent nearest neighbor couplings applied to a flat spin chain to capture the curved background. Different colors on the line plot show how our site-dependent coupling changes according to different $\ell_{\rm max}$ values. b,c) Checks for scrambling and chaos in the hyperbolic Ising model, (b) using tensor networks for OTOCs, and (c) using Krylov subspace methods.}\label{fig:summary}
	}
\end{figure*}
Quantum many-body chaos and scrambling have attracted a great deal of attention in recent years, from more theoretical endeavors like studying the scrambling dynamics of black holes using their dual descriptions~\cite{kobrin_many-body_2021-1,PhysRevB.94.035135,asaduzzaman_sachdev-ye-kitaev_2024,brown_quantum_2023,schuster_many-body_2022,cotler_black_2017,joshi_probing_2022} to more practical applications like investigating scrambling in chemical reactions to quantum cryptography~\cite{doi:10.1073/pnas.2321668121,chamon2024fast} and many more\cite{landsman2019verified,schuster2022many,blok2021quantum}. 

In almost all of these studies, the Sachdev-Ye-Kitaev (SYK) model has been the center of attention as the quintessential example of a model that exhibits fast scrambling and quantum chaos~\cite{PhysRevLett.70.3339,noauthor_alexei_nodate,noauthor_alexei_nodate2,maldacena_remarks_2016}. However, due to the all-to-all and random interactions in this model, any kind of classical and quantum simulation can be very challenging and often requires various truncations and approximations~\cite{xu_sparse_2020,garcia-garcia_sparse_2021,CAO20201170,sunderhauf_quantum_2019}. In this  paper, we advocate for a new model with only local site-dependent interactions that is also maximally chaotic and exhibits fast scrambling.~\footnote{There are other examples of fast scrambling models besides the SYK model, but to the best of our knowledge, they all require longer-range interactions than nearest neighbor interactions to be present in the model~\cite{PhysRevLett.126.200603,PhysRevResearch.2.043399,lashkari_towards_2013,PhysRevLett.125.130601,PhysRevA.102.022402,bentsen2019fast,chen2019quantum}.} 

A wide variety of tests have been developed 
to diagnose and classify quantum chaos,
including spectral form factors \cite{PhysRevX.8.021062,PhysRevD.98.086026,PhysRevLett.121.264101}, level statistics \cite{berry1977level,bohigas1984characterization}, different definitions of complexity\cite{roberts2017chaos,balasubramanian2022quantum}, Loschmidt echos \cite{peres1984stability,jalabert2001environment,gorin2006dynamics} and out-of-time ordered correlators (OTOCs)\cite{sekino_fast_2008,Maldacena:2015waa,swingle2016measuring,xu_accessing_2020,swingle2018unscrambling,xu2024scrambling}. Out of these many tools, Krylov Complexity --- which can be defined for both states and operators --- and OTOCs have emerged as the most comprehensive tests that have become the primary norm for diagnosing quantum chaos in many body systems~\cite{parker_universal_2019,rabinovici_krylov_2022,nandy_quantum_2024,baggioli_krylov_2024,chapman_krylov_2024,Jha:2024nbl,alishahiha_krylov_2024, Swingle:2018ekw,PhysRevB.97.144304,Garcia-Mata:2022voo, 10.21468/SciPostPhys.7.2.022,balasubramanian2025chaos}. 

While OTOCs have become the leading test in classifying quantum chaos and scrambling, their calculation is cumbersome and the numerical analysis of extracting Lyapunov exponents from OTOCs requires carefully fitting an exponential in a narrow parameter window. Also, it is possible for integrable models to display fast scrambling behavior in the absence of chaos~\cite{dowling2023scrambling,xu2020does,hashimoto2020exponential,pappalardi2018scrambling}.  On the other hand, we have diagnostic tools that are based on Krylov subspace methods, which do not require a fitting process as delicate as the OTOCs; however, due to the numerical instabilities in the Lanczos algorithm, they require either a full or partial re-orthogonalization procedure, limiting the system sizes that can be reached with these methods~\cite{Sanchez-Garrido:2024pcy}. Here, we use both of these techniques in addition to exact diagonalization to inspect the level statistics for providing a complete picture that encompasses large system sizes and a concrete diagnostic of quantum chaos and scrambling.

 A summary of our results can be seen in Fig.~\ref{fig:summary}, where a simple model inspired by holography --- the Hyperbolic Ising model --- satisfies multiple criteria of quantum chaos and fast scrambling. 
 We evaluate the scrambling time $t_s$, which  we define as the time it takes for the information initialized in the middle of spin chain (lattice) to reach the boundary (end points of the lattice), and find a logarithmic lightcone, with $t_s \sim \log(N)$, where
 $N$ is the number of spins. The Lyapunov exponents obtained from OTOCs at finite temperatures decay as $a/\beta$, where $\beta=1/T$ is the inverse temperature and $a$ is a constant. For a certain set of parameters, the effective Lyapunov exponents approach the scale set by the Maldacena-Shenker-Stanford (MSS) bound within numerical uncertainty.~\cite{Maldacena:2015waa}. Next, we study the Krylov state complexity (K-complexity), which exhibits the expected ramp-peak-plateau behavior \cite{Jha:2024nbl,alishahiha_krylov_2024}, the Krylov Entropy (K-entropy), which rises exponentially at early times, and the Krylov operator complexity, which shows the expected exponential-linear-saturation behavior~\cite{parker_universal_2019,rabinovici_krylov_2022,nandy_quantum_2024}. Finally, we study the spectral statistics of this model, which exhibits the expected Gaussian orthogonal ensemble (GOE) distribution in a chaotic model; all of these are indicative of fast scrambling and chaos.

These tests provide strong evidence that this model exhibits quantum chaos and can be classified as a fast scrambler, giving us one of the few examples where a fast scrambling behavior has been seen outside of the SYK model and the only example with computational resources needed for its classical and quantum simulations, similar to the mixed field Ising model. This discovery opens new avenues of research where we can use locally varying interactions to control different behaviors of scrambling and information propagation in quantum systems and test quantum chaos and scrambling without incurring the potentially high computational costs of an all-to-all connected model.

\section{Hyperbolic Ising Model }
Our model of interest, the Hyperbolic Ising model, is a mixed field Ising (MFI) model formulated on a one-dimensional hyperbolic space. In Fig.~\ref{fig:summary}, we show how one can visualize this model as a line (geodesic) between two boundary points in $AdS_3$ and interpret the effects of the background curvature as site-dependent couplings.  

The Hamiltonian that describes this Ising chain can be given as~\cite{PhysRevD.109.054513,ueda_hyperbolic_2011,ueda_transverse_2010,ueda_scaling_2010,Brower:2022atv}, 
\begin{align}
    \hat{H} =& -J\sum_{i}\left(\frac{\eta_i+\eta_{i+1}}{2}\right)\sigma_i^z\sigma_{i+1}^z \label{hamiltonian} \\ 
    &+h\sum_i \eta_i\sigma_i^x 
    +m\sum_i \eta_i\sigma_i^z, \nonumber \\ 
    &\text{with}\hspace{0.2cm}   \eta_i=\cosh({\rho_i}) \nonumber   
\end{align} 
where $\sigma^p_i$ is a local Pauli operator at site $i$ with $p \in \{x,y,z\}$. The local site-dependent coupling terms for the Ising chain, $\eta_i=\cosh({\rho_i})\sim \sqrt{g}$ arise from the Euclidean metric of $\mathrm{AdS}_2$,
\begin{align}
    \label{ads2}
        ds^2 & = \; \ell^2(\cosh^2(\rho)dt^2+d\rho^2)
\end{align}
To obtain the couplings in Eq.~\ref{hamiltonian} we set $\ell=1.0$ and discretize radius of curvature $\rho$ as follows, 
\begin{align}
        \rho_i & = \; -\ell_{\rm max}+i\frac{2\ell_{\rm max}}{N-1}
\end{align}
here $N$ is the size of the spin chain and $\ell_{\rm max}$ controls the curvature of the underlying curved space. One can easily recover the flat mixed field Ising model by taking the  $\ell_{\rm max} \to 0$ limit. This model exhibits a phase transition~\footnote{We use open boundary conditions in the simulations.} at $J/h = 1.0$ for a small value of the parameter $m$ similar to the flat Ising model~\cite{PhysRevD.109.054513}.

In the rest of the paper, we will center our discussion around this critical point where the mixed field Ising model is known to be chaotic~\cite{PhysRevLett.106.050405} and show that the inclusion of the curved background enhances this chaotic behavior and introduces fast scrambling. 

\begin{figure}[!h]
	\centering
	{
		\includegraphics[scale=0.4]{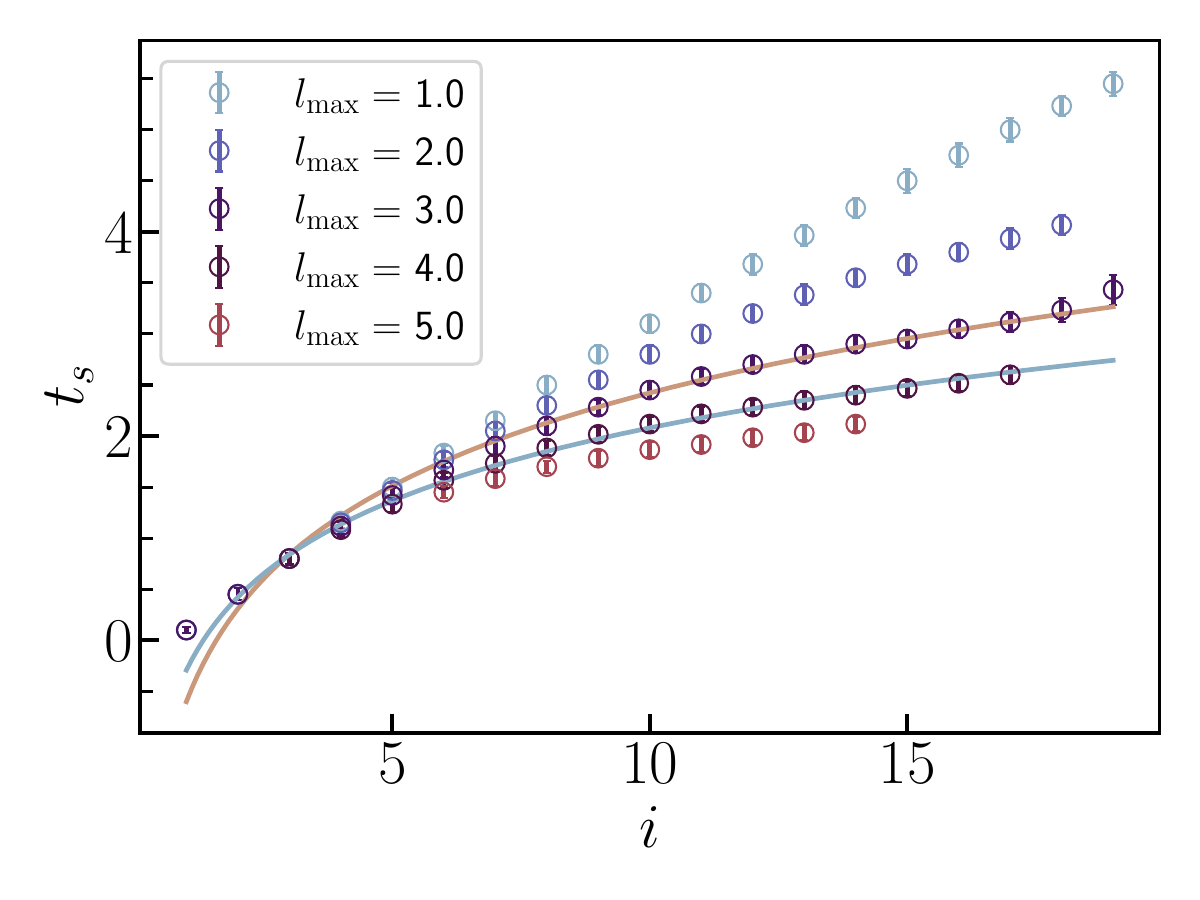}
		\caption{Scrambling time $t_s$ as a function of site index $i$ for $N=37$, $\ell_{\rm max}=\{1.0,...5.0\}$ at infinite temperature, showing the lightcone changing from linear to logarithmic as a function of $\ell_{\rm max}$.}
        \label{fig:otoc_inf}
	}
\end{figure}

\begin{figure*}
	\centering
	{
		\includegraphics[width=\textwidth]{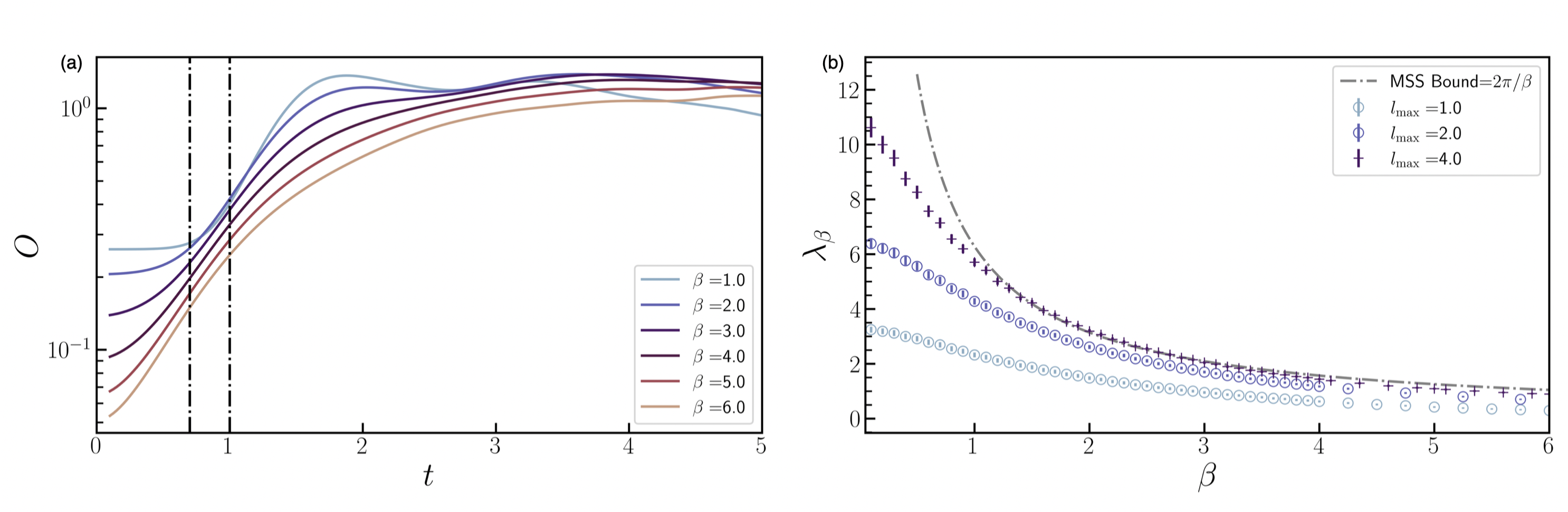}
		\caption{a) $O$ for $N=13$, $J/h=1.0$, $m=0.05$, and $\ell_{\rm max}=4.0$ black dashed lines show the fitting range we used in extracting the Lyapunov exponents. b) The Lyapunov exponent $\lambda_{\beta}$; for $\ell_{\rm max}=\{1.0,2.0,4.0\}$. The effective $\lambda_\beta$ approaches the MSS scale as the background curvature is increased, within the finite-size and fitting-window uncertainty.
        \label{fig:lyap}}
	}
\end{figure*}

In an earlier work~\cite{PhysRevD.109.054513}, we investigated information propagation and scrambling in this model in the infinite temperature limit, and showed that for a suitable subset of the parameters the lightcone obtained from infinite temperature OTOCs is logarithmic such that the saturation time $t_s \sim \log(i)$ where $i$ is the site index which implies that  $t_s \sim \log(N)$ for a system of $N$ spins. This logarithmic spreading can be seen in Fig.~\ref{fig:otoc_inf}, where lines show fits to a logarithmic function, and it is one of the tell-tale signs of fast scrambling~\cite{sekino_fast_2008}.  Surprisingly, this simple model with local interactions captures this behavior quite easily due to its unique site-dependent couplings arising from the curved background, contrary to the usual expectations that require all-to-all or long-range interactions to achieve such behavior. 

\section{OTOCs at Finite Temperature}

OTOCs are one of the most heavily used diagnostic tools for scrambling and chaos. They are obtained from the double commutator of operators $W(t)=e^{iHt}We^{-iHt}$ and $V$ and can be given as
\begin{align}
    F_{ij}(t) =& \langle \lvert \lvert [W_i(t), V_j] \rvert\rvert^2\rangle \\
    &\sim \frac{1}{Z} \Tr\left[ e^{-\beta H}W^\dagger_i(t) V^\dagger_j W_i(t) V_j\right] \nonumber
\end{align}
where $Z=\Tr(e^{-\beta H})$ and $\beta=1/T$. 

For a chaotic system, OTOCs are expected to show exponential growth as $F_{\beta}(t)\sim e^{\lambda_\beta t}$ for times before scrambling time, where $\lambda_\beta$ is the Lyapunov exponent and can be thought of as the quantum counterpart of the classical Lyapunov exponent.  
At finite temperatures, it is common to employ a regularized definition for the OTOCs due to divergences~\cite{Maldacena:2015waa}. This regularized form for the OTOCs  can be defined as
\begin{equation}
    \Tilde{F}_{ij}(t) = \Tr\left[ \alpha W^\dagger_i(t) \alpha V^\dagger_j \alpha W_i(t) \alpha V_j\right],
\end{equation}
where $\alpha^4=Z^{-1}e^{-\beta H}$. 

$W_i$, and $V_j$ are $\sigma_{\frac{N+1}{2}}^z$ and $\sigma_1^z$ respectively. We obtain the finite temperature states that go into this calculation by purification techniques\cite{PhysRevB.72.220401,PRXQuantum.2.040331,feiguin2005finite} developed for Matrix Product States (MPS)~\cite{white2004real,PhysRevB.48.10345,RevModPhys.77.259,PhysRevLett.91.147902,verstraete2004matrix,PhysRevLett.93.040502}.~\footnote{For tensor network simulations we used the ITensor Library~\cite{Fishman:2020gel}; details of our simulation procedure can be found in Appendix.~\ref{appx:thermal_state_prep}.}
After the desired thermal state is obtained, it is then time evolved while measuring  $O=1- \Tilde{F}_{ij}(t)$ at each time-step. We begin our OTOC studies by choosing model parameters the same as preliminary results obtained from analyzing scrambling time behavior as a function of the system size as shown in Fig.~\ref{fig:otoc_inf}, i.e. $J/h=1.0$, and $m=0.05$ In Fig.~\ref{fig:lyap} (a) we show our results for the OTOCs at $\ell_{\rm max}=4.0$ at inverse temperatures ranging from $\beta=[0.5,...6.0]$. To extract the Lyapunov exponents, we fit $\log(\frac{dO}{dt})$ to $f(x)=at+b$; this approach yields better fits with better $R^2$ results compared to exponential fits of the original function~\cite{shen_out--time-order_2017}.

Our results presented  in Fig.~\ref{fig:lyap} 
 (a) identifies a regime consistent with the expected exponential rise of the OTOCs with the two dashed lines indicating the fitting range which is short-lived due to finite system sizes.  This range is chosen such that for all the values of $\beta$, the linear fits result in a $R^2\ge0.96$. 

Fig.~\ref{fig:lyap} (b) shows the decay of the $\lambda_\beta$ as $a/\beta$ where $a$ is some constant. We observe a behavior consistent with the Maldacena-Shenker-Stanford (MSS) bound on chaos (often considered as a criterion for fast scrambling) for Lyapunov exponents for small temperatures as the background curvature is increased ($a \to 2\pi$). However, the saturation constant $a$ which is obtained from fitting the data, is highly sensitive to the range of OTOC data chosen to fit an exponential function.

Our analysis so far relies on extracting Lyapunov exponents from exponential fits to the OTOCs, which necessarily involves selecting a relatively narrow time window. In large systems, such exponential behavior is expected to persist over extended time scales; however, in our case the accessible system sizes are limited by the growth of bond dimension in tensor network simulations. As a result, any exponential regime in the OTOCs is short-lived and does not exhibit a clear parametrically long window. Consequently, the extracted Lyapunov exponents should be interpreted as effective, finite-size quantities rather than asymptotic indicators of chaos. Further observations crucial for understanding this OTOC behavior were obtained using an alternative data fitting scheme as shown in Appendix~\ref{appx:otocs} and these also show the expected $a/\beta$ decay.

\begin{figure}
    \centering
    \includegraphics[width=0.5\textwidth]{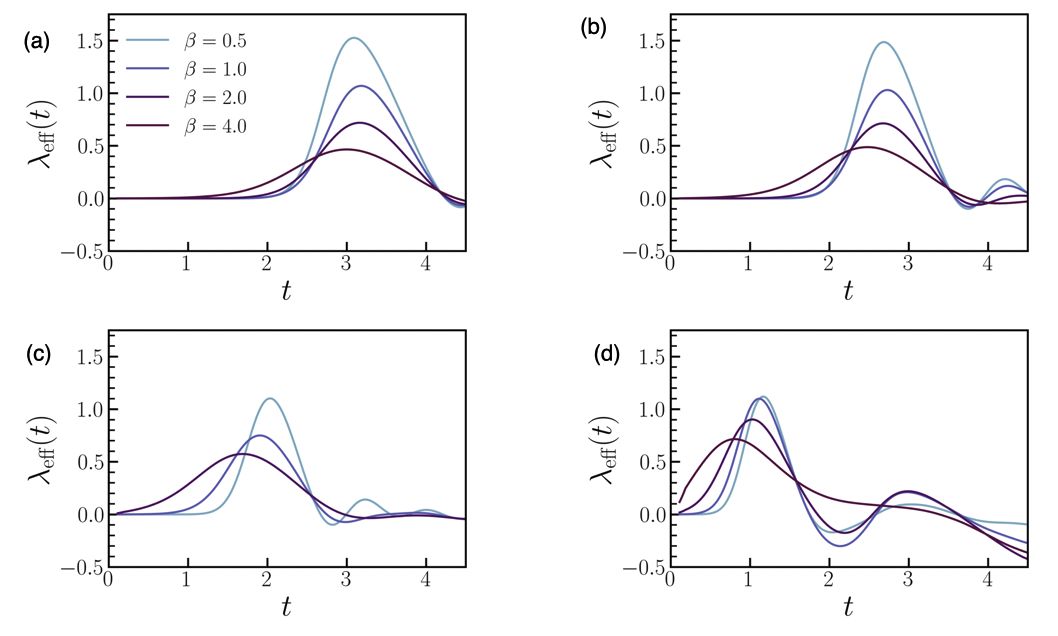}
    \caption{$\lambda_{\rm eff}(t)$ at $\beta = \{0.5,1.0,2.0\}$ for $\ell_{\rm max}=\{0.0,1.0,2.0,4.0\}$}
    \label{fig:lambda_eff}
\end{figure}

To further probe the scrambling dynamics in a manner less sensitive to fitting windows, we also compute the instantaneous growth rate $\lambda_{\rm eff} := \tfrac{1}{2}\frac{d}{dt}\ln O(t)$. In large systems, sustained exponential growth would manifest as a plateau in $\lambda_{\rm eff}(t)$; however, for the system sizes considered here, such plateaus are not clearly developed. Instead, we use $\lambda_{\rm eff}(t)$ to characterize the transient growth regime and to examine its dependence on the background curvature, as shown in Fig.~\ref{fig:lambda_eff}.

\begin{figure}
    \centering
    \includegraphics[width=0.45\textwidth]{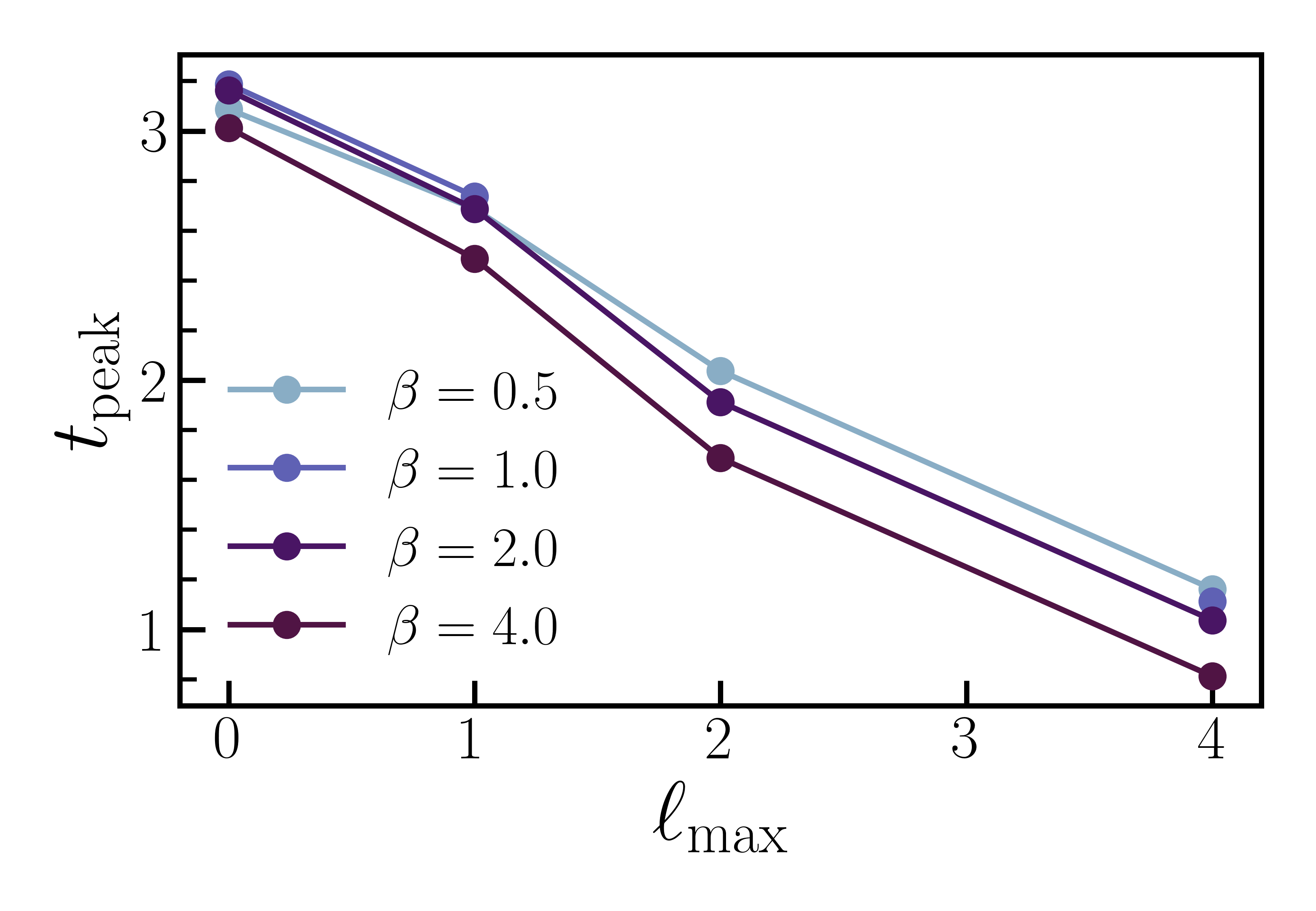}
    \caption{Dependence of $t_{\rm peak}$ on curvature $\ell_{\rm max}$}
    \label{fig:tpeak}
\end{figure}
As anticipated from the finite system sizes considered, $\lambda_{\rm eff}(t)$ does not exhibit a well-defined plateau, reflecting the absence of a sustained exponential growth regime. Instead, $\lambda_{\rm eff}(t)$ displays a pronounced peak at a characteristic time $t_{\rm peak}$, which serves as a useful proxy for the onset and duration of the transient growth regime. In Fig.~\ref{fig:tpeak}, we plot $t_{\rm peak}$ as a function of $\ell_{\rm max}$ and observe a clear decrease in $t_{\rm peak}$ with increasing curvature supporting our claim that an increase in background curvature decreases the scrambling time.  While this quantity should be interpreted as a finite-size diagnostic rather than an asymptotic scrambling time, its systematic dependence on $\ell_{\rm max}$ provides evidence that increasing curvature enhances the rate of information scrambling in the system.

 From our investigations of OTOCs, we see that OTOCs alone are insufficient to classify this model as a fast scrambler at the system sizes of this study, even though we see some agreement with the MSS bound. Larger studies are needed to fully demonstrate that this model saturates the MSS bound with long-lasting exponential behavior in OTOCs. Moreover, large inverse temperature behavior studies must take into account the Trotter errors due to Trotterized evolution used in both obtaining the thermal state and the time-evolved states. Hence, the range of $\beta$ values that can be reliably simulated are limited which affects the fitting window to obtain the saturation constant.
 
 To provide more observables to classify this model as a chaotic fast scrambler, we further inspect Krylov complexity and spectral statistics in addition to OTOC studies. Only when all these observables agree on the scrambling behavior do we classify the model as fast scrambling.
\begin{figure*}
{
    \centering

    \includegraphics[width=\textwidth]{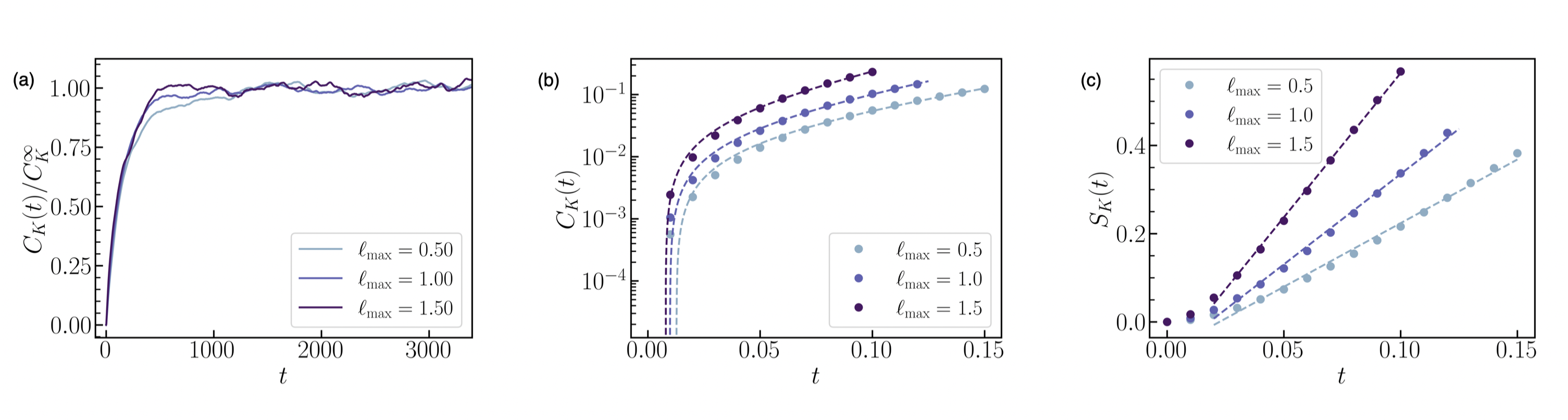}
    \caption{a) $C_K$ using operators  for $N = 6, J/h = 1.0, m = 0.25$ at different $\ell_{\rm max}$ values. b) Exponential fit of $C_K$ at early time scales for the same set of parameters shown on a log-linear scale. c)Linear fit of $S_K$ at early time scales for the same set of parameters.}
    \label{fig:ocomplexity}
}
\end{figure*}

\section{Krylov Observables}

\begin{figure}
{
    \centering
    \includegraphics[width=0.45\textwidth]{figs/Lancozs_N6_m0.25_Jbh_1.0.png}
	\caption{Lanczos coefficients $b_n$ as a function of the krylov state index $n$ for $N = 6, J/h = 1.0, m = 0.25$.}
    \label{fig:operator_lancoz}
}
\end{figure}
\begin{figure} \label{fig:scomplexity}
	{
		\centering
		\includegraphics[width=0.45\textwidth]{figs/hyperbolic_ising_N11_J1.0_h1.05_m0.25_staggered_state.png}
		\caption{State Complexity for $N = 11, J/h = 1.0, m = 0.25$. The inset shows the change in peak of the K-complexity with increase in the background curvature, $\ell_{\rm max}$.}
		
	}
\end{figure}
In this section we discuss observables obtained using Krylov subspace methods. The main advantage of these observables is that they do not require a delicate fitting scheme like OTOCs do to diagnose chaos. 
The two quantities of interest are the state and operator version of the K-complexity $C_{K}$ and Krylov Entropy $S_{K}(t)$ calculated using operators.  These observables obtained from the Krylov basis wavefunctions  that exhibit  different characteristics for integrable and chaotic systems allowing them to be used as a diagnostic tool for quantum chaos~\cite{parker_universal_2019,rabinovici_krylov_2022,nandy_quantum_2024,baggioli_krylov_2024,chapman_krylov_2024,Jha:2024nbl}. The Krylov state complexity $C_{K}(t)$ and Krylov Entropy $S_{K}(t)$ for operators can be defined as, 
\begin{align}
    C_{K}(t) & = \; \sum_{n=0}^{\mathbb{K}-1} n\lvert\phi_n(t)\rvert^2\\
    S_{K}(t) & = \; -\sum_{n=0}^{\mathbb{K}-1} \lvert\varphi_n(t)\rvert^2 \ln{\left(\lvert\varphi_n(t)\rvert^2\right)}
\end{align}
where $\phi_n(t), \varphi_n(t)$ are the Krylov basis wavefunctions obtained via Lanczos algorithm~\cite{lanczos1950iteration} for states and operators respectively with $\mathbb{K} = dim(K)$, is the dimension of the Krylov subspace $K$. 

The dynamics in the Krylov subspace are fully determined by the time evolution of Krylov wavefunctions and due to the tridiagonal structure of the Lanczos coefficients time evolution in the Krylov subspace simplifies to the following recursion relation~\cite{Sanchez-Garrido:2024pcy}.

\begin{equation}
    \label{eq:rec_op}
    \dot{\phi}_n(t) = b_n\phi_{n-1}(t) -b_{n+1}\phi_{n+1}(t) 
\end{equation}

Here  $\{(a_n, b_n)\}_{n=0}^{\mathbb{K}-1}$  are the Lanczos coefficients obtained via the Lanczos algorithm~\cite{lanczos1950iteration} and the equations can be solved with the initial condition $\phi_n(t=0) = \delta_{n0}$.

Furthermore, these recursion relations map the time evolution in the full space to a single particle hopping model in the one-dimensional Krylov chain making long-time simulations tractable, especially for time scales where our tensor network methods break down. More details about the Lanczos algorithm and our implementation can be found in the Appendix.~\ref{appx:krylov} 

\begin{figure*}
{
    \centering
    \includegraphics[width=\textwidth,height=0.3\textwidth]{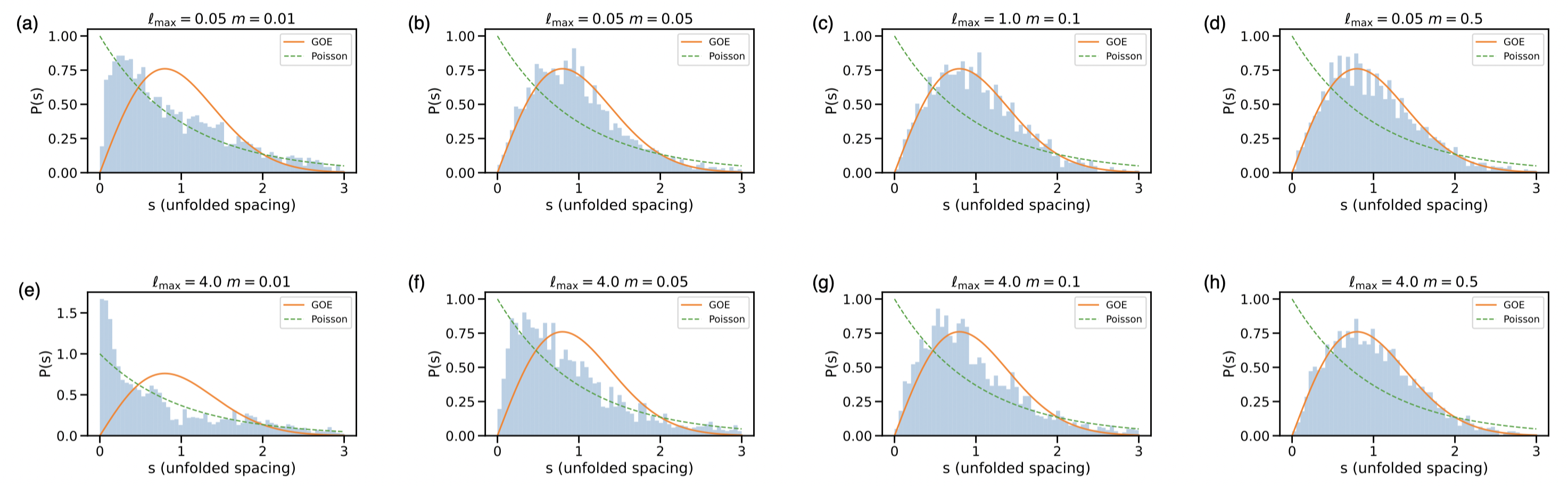}
	\caption{Spectral statistics for $N=13$, $J=1.0$, $h=1.05$, $m=\{0.01,0.05,0.1,0.5\}$ with (a)-(d) at $\ell_{\rm max}=0.05$ and (e)-(f) at $\ell_{\rm max}=4.0$.}
    \label{fig:GOEstats}
}
\end{figure*}  

The first set of observables are based on calculations using operators. Lanczos algorithm can be generalized to operators in straightforward way using the channel-state duality~\cite{PhysRevA.87.022310}.  In this case, the Hamiltonian becomes a Liouvillian that acts on the operators, and the usual inner product $\langle p \lvert q\rangle$ of quantum states in Lanczos Algorithm is changed to $\left( A \lvert B \right):= \frac{1}{D} Tr\left[A^\dagger B\right]$. We choose $\sigma^z_1$ as our initial seed operator for the Lanczos algorithm and carry out the subspace generation and time-evolution. 

The $C_K(t)$ obtained using the operators is closely related to operator growth in the full model and hence can be used to identify scrambling behavior and chaos~\cite{nandy_quantum_2024}. For a fast scrambler, operator complexity is expected to grow exponentially at early timescales followed by a linear increase and finally, they saturate/oscillate about some fixed average value related to the dimension of the Krylov subspace. $S_k(t)$ on the other hand is expected to show linear rise when $C_K(t)$ is exponentially increasing followed by logarithmic increase during the linear growth of the complexity\cite{parker_universal_2019,rabinovici_krylov_2022,nandy_quantum_2024,baggioli_krylov_2024,chapman_krylov_2024,Jha:2024nbl}. In Fig.~\ref{fig:ocomplexity} (a) we show normalized operator complexity for different curvatures, where the normalization constant is the late time average of the complexity. In Fig.~\ref{fig:ocomplexity}(b) we show corresponding exponential fits at early time scales on a logarithmic y-axis, where we observe that the argument of the exponential increases with increase in $\ell_{\rm max}$, whereas in Fig.~\ref{fig:ocomplexity}(c) we show a linear fit to $S_K(t)$ at the similar early times scales. Also the Lanczos coefficients $b_n$ should have a linear relationship with $n$, where $0 \leq n \leq \mathbb{K}-1$. This shape that the  Lanczos coefficients display is generally referred to as the Lanczos descent~\cite{barbon_evolution_2019} which can be seen in Fig.~\ref{fig:operator_lancoz} where show the Lanczos coefficient $b_n$ vs $n$.

The second set of observables that signal a chaotic system in the context of Krylov methods is the state K-complexity. We start the Lanczos algorithm with the state $\ket{+^y}\otimes\lvert1\rangle\otimes\ket{+^y}\otimes\lvert1\rangle\ldots\otimes\ket{+^y}\otimes\lvert1\rangle$, where $\ket{+^y} = \frac{1}{\sqrt{2}}\left(\ket{0}^y + i\ket{1}^y\right)$. In this initial seed, $\lvert+^y\rangle$ is on every other lattice site starting with the first site, and $\lvert1\rangle$ is on every other lattice site starting with the second site. We chose this initial seed because in the flat Ising model, this state exhibited a clear peak in $C_K(t)$ in the chaotic regime of the model~\cite{alishahiha_krylov_2024}.\footnote{The choice of the initial seed is an essential part of this algorithm, and it can have an effect on the results obtained for $C_K(t)$ and $S_K(t)$. To alleviate this dependence on the initial seed, new algorithms that sample a wide array of initial seeds have been proposed~\cite{PhysRevLett.134.050402}.} Using this initial choice, we carry out the subspace generation and time evolution and investigate $C_K(t)$.  

The key indicator of quantum chaos obtained using states is that $C_K(t)$ is expected to exhibit ramp-peak-plateau behavior in time.\cite{parker_universal_2019,rabinovici_krylov_2022,nandy_quantum_2024,baggioli_krylov_2024,chapman_krylov_2024,Jha:2024nbl}. 

Our results confirm the expectations described above and show that the system becomes more chaotic as the background curvature increases.  In Fig.~\ref{fig:scomplexity} we show the normalized $C_K(t)$ where the ramp-peak-plateau behavior is clearly visible and the peak becomes more prominent as we increase $\ell_{\rm max}$ in line with the expectation that the model becomes more chaotic as $\ell_{\rm max}$ is increased. The normalization constant is again the late time average of the complexity.

Our results for the Krylov subspace methods which can be summarized numerically in Appendix.~\ref{appx:krylov_data} show the expected behavior, thus, providing evidence independent of OTOCs on fast scrambling characteristics of the Hyperbolic Ising model.

\section{Spectral Statistics and Emergence of Quantum Chaos}\label{appx:spectral}
Now, we provide further evidence that this model is chaotic by investigating its level repulsion spectra and showing that it follows GOE statistics as expected from a chaotic model~\cite{berry1977level,bohigas1984characterization,hsu1993level,poilblanc1993poisson,casati1980connection,d2016quantum,oganesyan2007localization}.
It is well known that to reliably resolve the statistics in the level repulsion spectrum, one needs to break all the symmetries of the model and focus on a well-defined sector of the Hilbert space. We achieve this by firstly breaking the reflection symmetry of the coupling by taking only the exponentially increasing part into consideration for the full system. After this, we use exact diagonalization and project to a fixed parity sector to resolve the rest of the symmetries.

\begin{figure}
{
    \centering
    \includegraphics[width=0.5\textwidth]{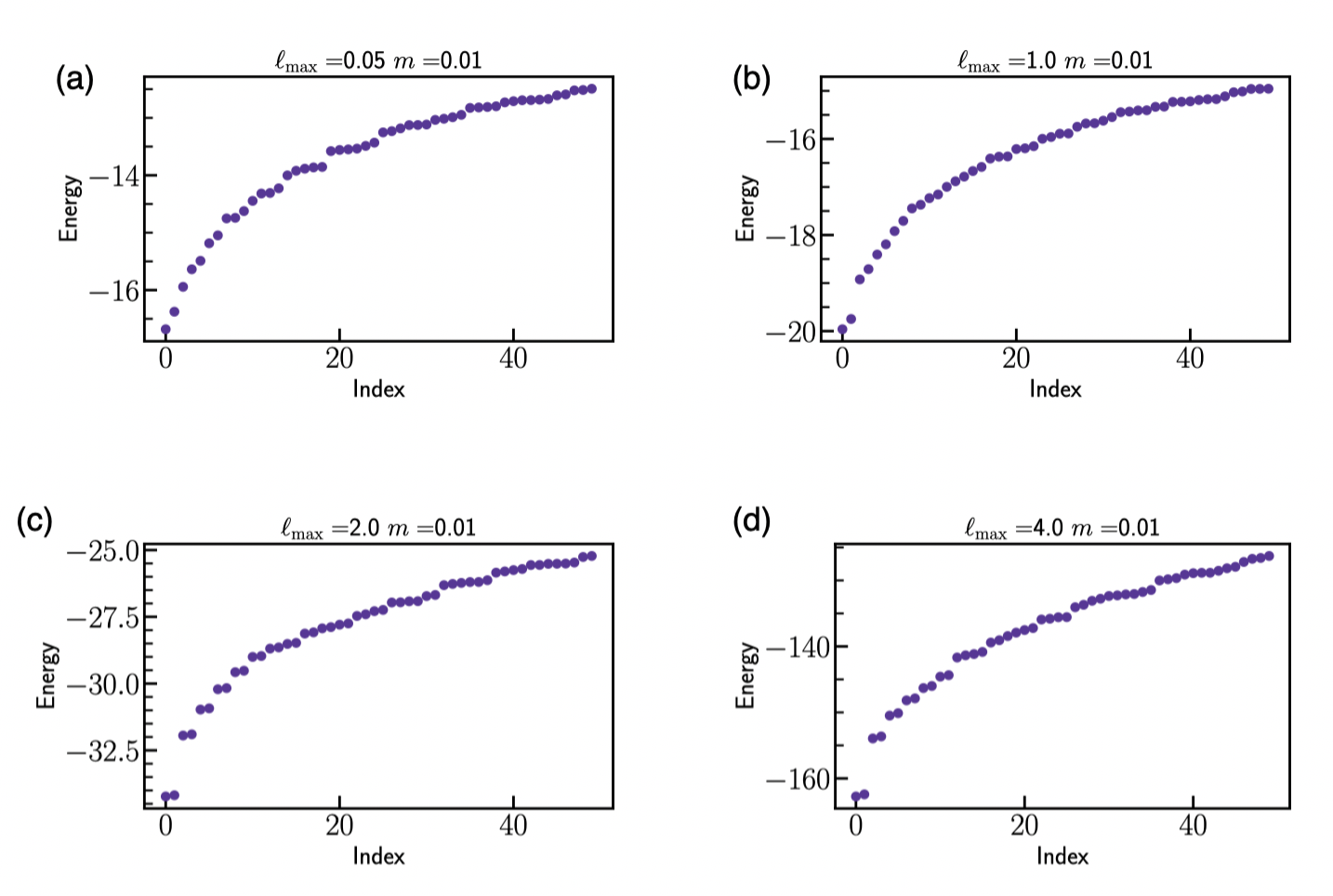}
	\caption{Low lying eigenvalues for  $N=13$, $J=1.0$, $h=1.05$, $m=0.01$ with $\ell_{\rm max}=\{0.05,1.0,2.0,4.0\}$ showing how the spectrum becomes degenerate as $\ell_{\rm max}$ increases.}
    \label{fig:eigenvalues}
}
\end{figure}

\begin{figure}
    \centering
    \includegraphics[width=0.45\textwidth]{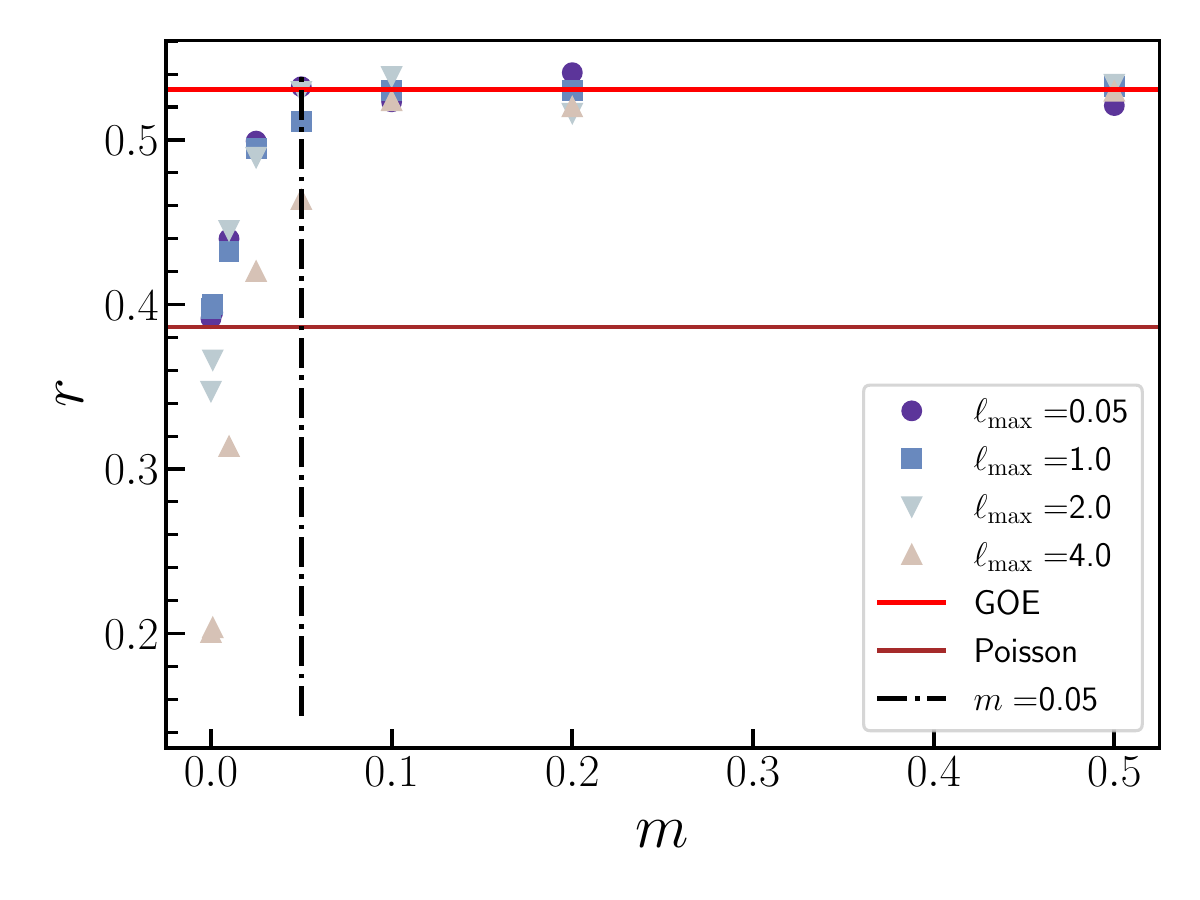}
    \caption{Level spacing ratio $r$ as  function of $m$ at $N=13$, $J=1.0$, $h=1.05$ and $\ell_{\rm max}=\{0.05,1.0,2.0,4.0\}$. The dashed line shows the choice of $m$ in the main text.}
    \label{fig:spectral_stats_vs_l}
\end{figure}

In Fig.~\ref{fig:GOEstats}, (a)-(d)  we show the level statistics at $\ell_{\rm max}=0.05$ where the model is very close to the flat mixed field Ising limit. Similarly Fig.~\ref{fig:GOEstats} (e)-(f) shows the level statistics at $\ell_{\rm max}=4.0$. From these plots, we can see how our system goes from a Poisson distribution to a GOE distribution as we increase the mass. We see that when the curvature is present, we need a higher $m$ value to obtain the GOE ensemble compared to the flat limit. The reason for this can be understood by inspecting the spectrum of the model. In  Fig.~\ref{fig:eigenvalues} (a)-(d), we plot $50$ low-lying eigenvalues of our model at $N=13$, $J/h=1.05$, $m=0.01$ as we vary the background curvature. We clearly see that as we increase the background curvature, the spectrum becomes more degenerate, especially for the lowest lying eigenstates like the ground state and the first couple of excited eigenstates. This results in a need for a larger $m$ value to break the degeneracy, which implies that there is an interplay between $\ell_{\rm max}$ and $m$ in observing chaotic spectral statistics.

This observation is further supported in Fig.~\ref{fig:spectral_stats_vs_l}, where we show how the average level spacing ratio $r$ approach the expected $r_{\rm GOE}$ $\sim 0.53$~\cite{Atas_2013} compared to the $r_{\rm Poisson}$ $\sim 0.38$~\cite{oganesyan2007localization} value as a function of $m$. This $r$ is defined as~\cite{Atas_2013}
\begin{equation}
    r = \langle r_n\rangle  = \frac{s_n}{s_{n-1}} \;\;,\;\; s_n = e_n - e_{n-1}
\end{equation}
where, $e_n$ is the $n^{\rm th}$ eigenvalue in the sorted set of eigenvalues for the Hamiltonian. The plot in Fig.~\ref{fig:spectral_stats_vs_l} clearly shows that one needs larger $m$ values to resolve the GOE behavior as $\ell_{\rm max}$ is increased and the choice of $m$ in the main text is clearly away from the Poisson value.

From the spectral statistics and level repulsion analysis we see even when the OTOCs classify the model as a fast scrambler the model could be close to the non chaotic limit depending on the value of $m$ and to truly classify the model as a fast scrambler we need total agreement between all of our observables which only happens when $m,\ell_{\rm max} > 0.0$

\section{Conclusion}
In this paper, we investigated chaos and scrambling in the Hyperbolic Ising model which is a nearest neighbor mixed field Ising model with site-dependent couplings using tensor network and Krylov subspace methods and showed that the observables obtained from both of these independent techniques show signatures of fast scrambling and chaos. 

 Using tensor networks, we showed that, for a suitable subset of the parameter space we observed a logarithmic lightcone with scrambling time $t_s\sim log(N)$. Finite temperature OTOCs show a short-time exponential rise at early times similar to the operator complexity, with  Lyapunov exponents obtained from these exponentials decaying as $a/\beta$ while saturating the MSS bound for a certain set of parameters. We have also investigated the dependence of the peak shown in the instantaneous growth rate $\lambda_{\rm eff }$ wrt background curvature. We saw that increasing the background curvature decreases the peak time in line with our claims that background curvature causing faster scrambling.
 
 With the Krylov subspace methods, we showed that the observables obtained using states exhibit the following signatures of chaos: K-Complexity demonstrated the expected ramp-peak-plateau behavior with the peak scaling according to the background curvature $\ell_{\rm max}$. Next, K-Entropy showed the expected exponential rise at early times. Finally, we showed that Lanczos coefficients $b_n$s depend linearly on $n$ which is another criterion of chaos. Operator complexity showed an exponential rise similar to the OTOCs at early times followed by a linear increase, and saturation around a maximum at late times.  
 
 Due to the limited system sizes accessible in our finite-temperature tensor-network calculations, the exponential growth regime of the OTOCs is necessarily short-lived. Therefore, the Lyapunov exponents extracted here should be interpreted as effective finite-size quantities rather than asymptotic large-system exponents. Our conclusions are instead based on the consistency of these OTOC results with the independent evidence from spectral statistics and Krylov diagnostics. Which shows that the model exhibits the expected GOE behavior in its spectral statistics and average level repulsion, indicating that the model is chaotic for the appropriate set of parameters.

Combining these results, we see robust evidence that this model can be classified as a fast scrambler when all four of our checks agree. This provides us with a unique model in the class of fast scramblers where with only nearest-neighbor interactions and site-dependent couplings fast scrambling behavior can be achieved.

\section*{Acknowledgements}
We thank Raghav Govind Jha and Bharath Sambasivam for valuable discussions. 
AFK and GCT acknowledge financial support from the National
Science Foundation under award No. PHY-2325080: PIF: Software-Tailored Architecture for Quantum Co-Design. AS is supported by U.S. Department of Energy grant DE-SC0019139. We acknowledge the computing resources provided by North Carolina State University High Performance Computing Services Core Facility (RRID:${\rm SCR\_}022168$).

\begin{appendix}
\numberwithin{equation}{section}

\section{Details on Thermal State Preparation}
\label{appx:thermal_state_prep}

The purification procedure used to obtain the thermal states can be quickly summarized as follows.  

We start with an infinite temperature Thermofield double state $\ket{\rm{TFD}(\infty)}$  which can be initialized as
\begin{equation}
\ket{\rm{TFD}(\infty)} = \mathbf{1}/\sqrt{2}
\end{equation}
into a Matrix Product Operator(MPO) and evolved under imaginary time evolution with Time-Evolving Block Decimation(TEBD) techniques (using $e^{-\beta H}$, where $H$ is the Hyperbolic Ising Hamiltonian). 

If one continues the time evolution indefinitely this algorithm eventually reaches the ground state of the model. However, stopping at a desired $\beta$  allows us to obtain the thermal states of the Hyperbolic Ising model in terms of an MPO.

With access to the thermal states of the model, we can calculate the OTOCs constructed using two local operators $W_i(t)$ and $V_j$ where  $W(t)=e^{iHt}W(0)e^{-iHt}$, combined into the double commutator.
\begin{equation}
    C(t) = \langle ||[W_i(t),V_j]||^2 \rangle=2(1-\mathrm{Re}[F_{ij}(t)]), \label{double_comm}
\end{equation}
where $<.> = \frac{1}{Z}\Tr[e^{-\beta H}.]$ corresponds to the thermal expectation value at inverse temperature $\beta$ and $F_{ij}(t)$ is known as the out-of-time-ordered correlator (OTOC).  Other terms in the double commutator become irrelevant after a short time scale and hence can be omitted.  
\begin{equation}
    F_{ij}(t) = \langle W_i(t)^{\dagger}V_j(0)^{\dagger} W_i(t)V_j(0) \rangle.\label{F_def}
\end{equation}

In our calculations, we take $W(t)=\sigma^z(t)$, $V=\sigma^z$ 
and fix the position of the $W(t)$ operator at the center of the lattice chain which is the site $(N+1)/2$. we then place the operator $V$ at different lattice sites $i$ and measure $F_{ij}(t)$

\begin{figure}
{
    \centering
    \includegraphics[width=0.5\textwidth]{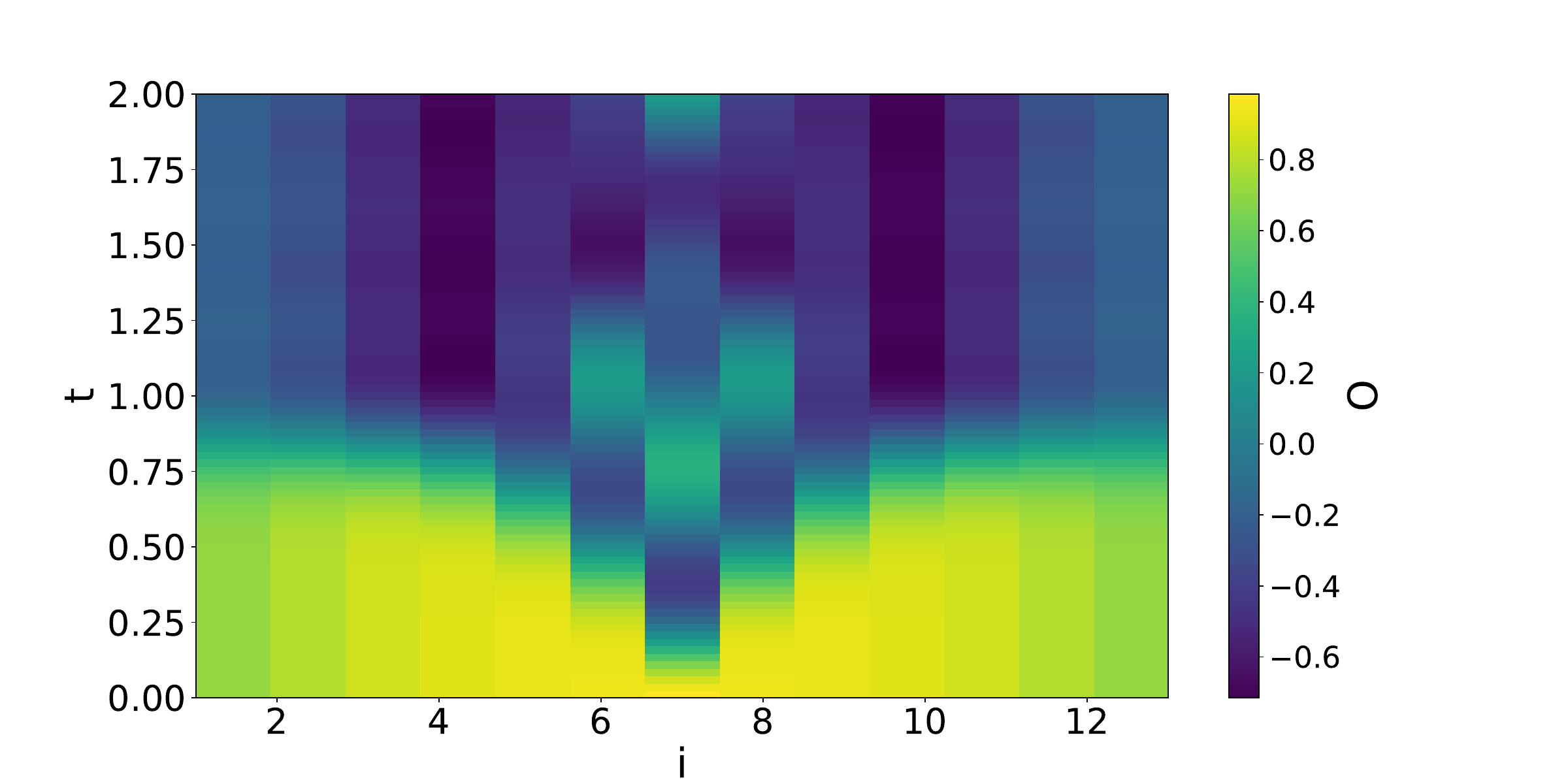}
	\caption{Lightcone at $N=13, J/h=1.0, m=0.05, \beta=0.25, \ell_{\rm max}=3.0$}
    \label{fig:otocl3}
}
\end{figure}

\begin{figure}
{
    \centering
    \includegraphics[width=0.5\textwidth]{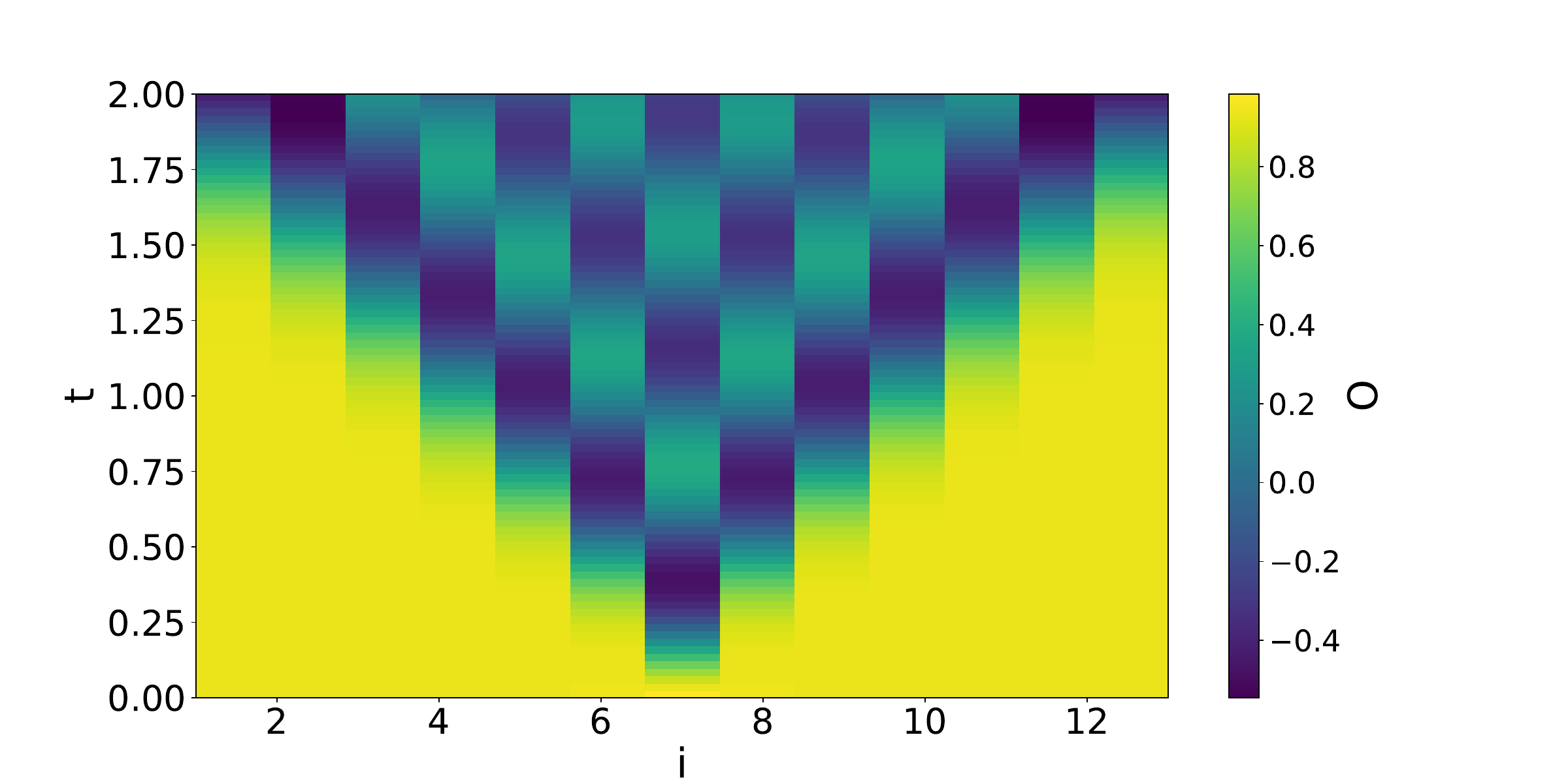}
	\caption{Lightcone at $N=13, J/h=1.0, m=0.05, \beta=0.25, \ell_{\rm max}=0.05$}
    \label{fig:otocl05}
}
\end{figure}

For constructing the lightcone we need to calculate the OTOC at all sites of the spin chain which leads to the following propagation patterns in Fig.~\ref{fig:otocl05},~\ref{fig:otocl3}. As can be seen from these two plots,  as the background curvature increases, the lightcone becomes warped. In \cite{PhysRevD.109.054513} we showed in detail that depending on the curvature $\ell_{\rm max}$ and the nearest-neighbor coupling strength $J$ it is possible to achieve different kinds of scrambling behavior ranging from linear to logarithmic using OTOCs calculated at infinite temperature.  

\section{Alternative fitting procedure for OTOCs}
\label{appx:otocs}
\begin{figure}
{
    \centering
    \includegraphics[width=0.5\textwidth]{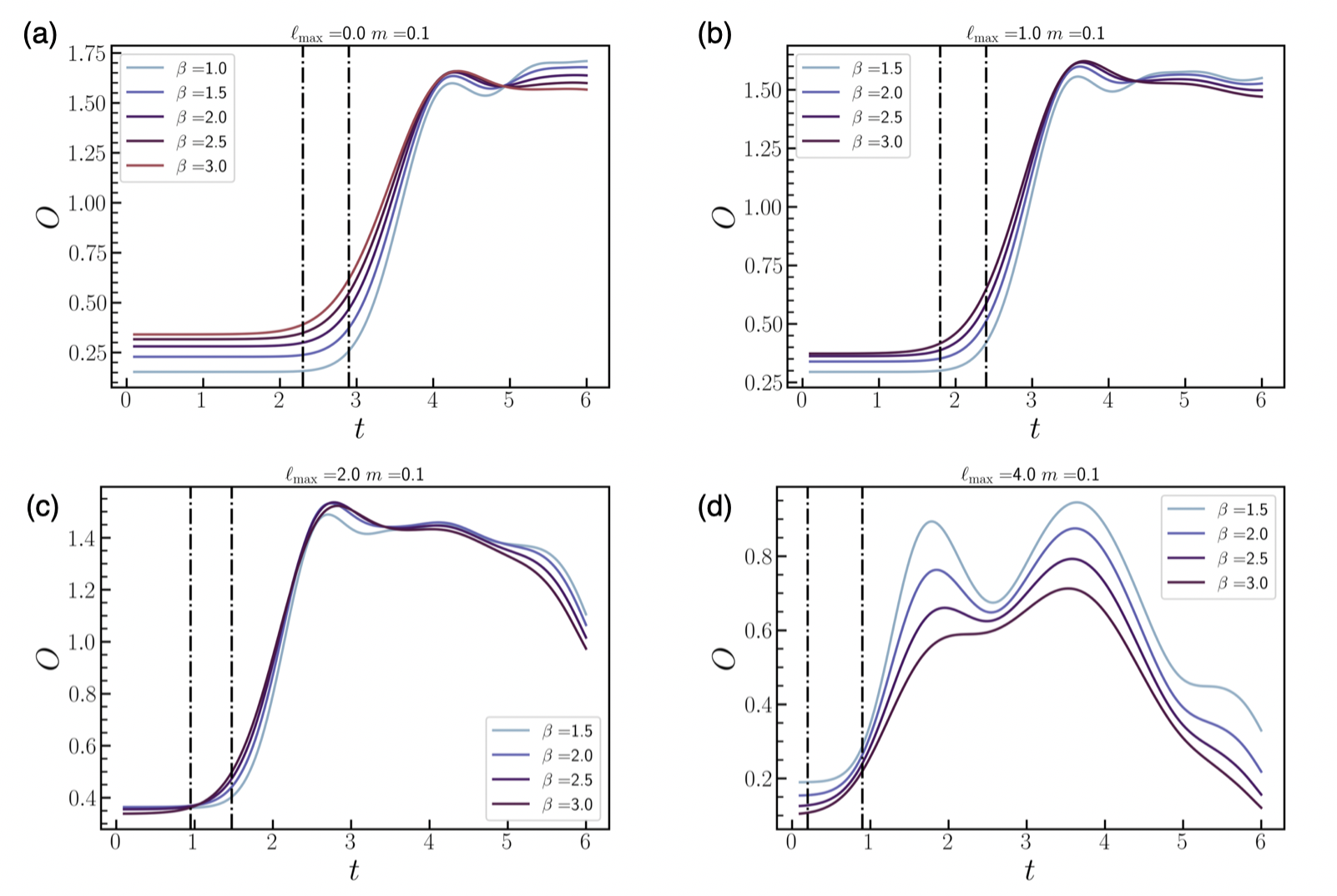}
	\caption{O for N=13, J/h=1.05 for $m={0.0,0.1}$ and $\ell_{\rm max}={0.0,1.0,2.0,4.0}$}
    \label{fig:otocs_m0.1}
}
\end{figure}

To further inspect the stability of the fitting procedure and the  relationship between $m$ and $\ell_{\rm max}$, we repeat a similar Lyapunov exponent analysis  for $m={0.0,0.1}$ and  $\ell_{\rm max}={0.0,1.0,2.0,4.0}$. To ensure that our fitting procedure was not skewing our results, for this analysis, we use a sliding window fitting procedure instead of a fixed fitting window, in which we specify an initial center and a range for values to be fitted to the left and to the right of the center as inputs. Then our fitting algorithm moves the center around and adjusts the window range to find the region with the best $R^2$ value.  In Fig.~\ref{fig:otocs_m0.1} we show the results for the OTOCs and the fitting range obtained from the sliding window fits for $m={0.0,0.1}$ and $\ell_{\rm max}=\{0.0,1.0,2.0,4.0\}$. It has been observed that a transient or consistent exponential behavior can be seen in integrable models~\cite{dowling2023scrambling,xu2020does,hashimoto2020exponential,pappalardi2018scrambling}, and we see that we can indeed obtain exponential fits in the $m=0.0, \ell_{\rm max}=0.0$ limit with good $R^2$ values.

\begin{figure}
{
    \centering
    \includegraphics[width=0.45\textwidth]{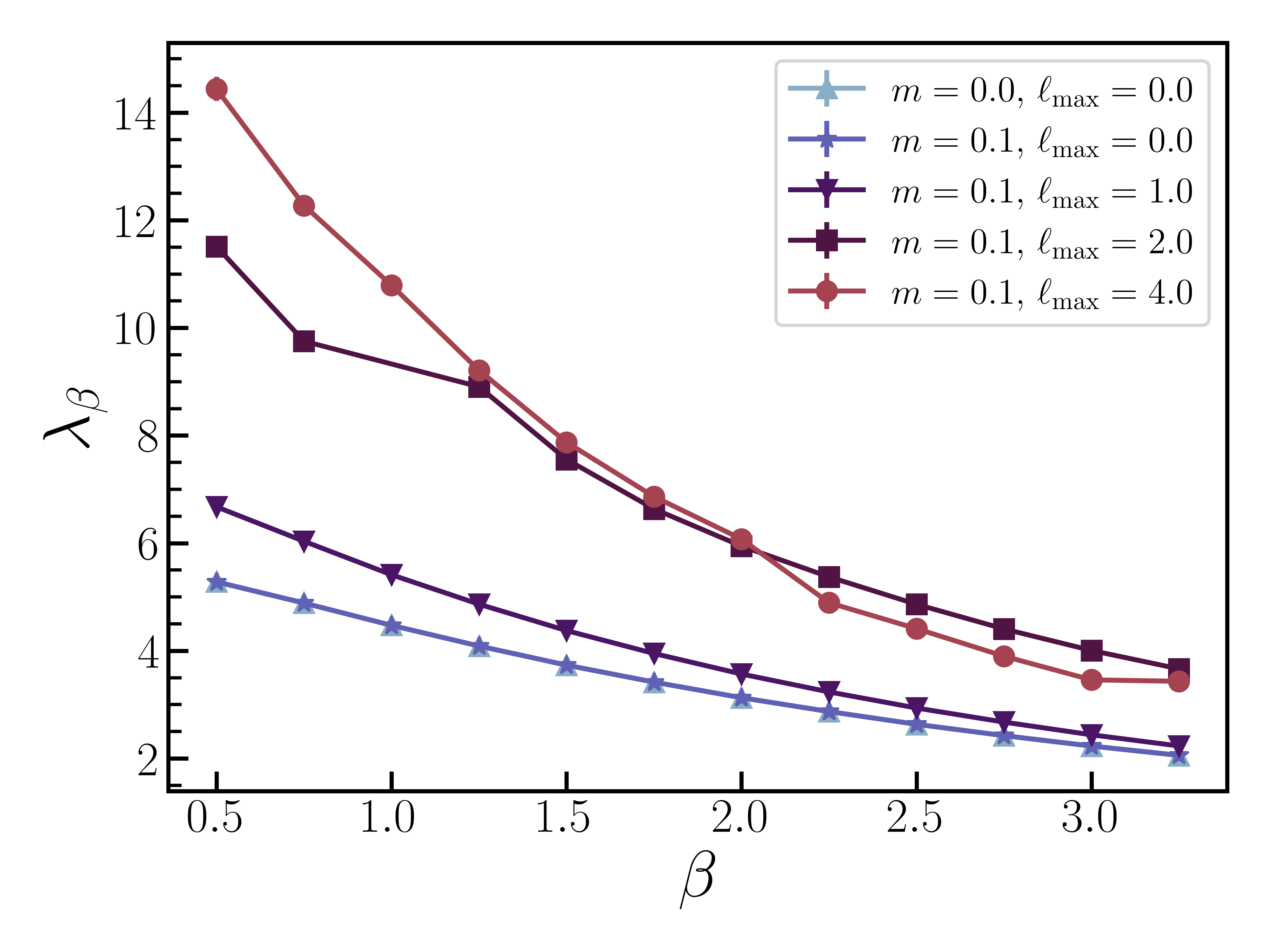}
	\caption{The Lyapunov exponent $\lambda_{\beta}$ obtained from the exponential fitting as a function of $\beta$}
    \label{fig:lyap_m_vs_l}
}
\end{figure}

However, in Fig.~\ref{fig:lyap_m_vs_l} we can clearly see that for the flat integrable limit $m=0.0$, $\ell_{\rm max}=0.0$ and flat non-integrable limit $m=0.1$, $\ell_{\rm max}=0.0$ we do not see a $1/\beta$ dependence on the Lyapunov exponents obtained from the exponential fits. Hence, for these parameters, the model can't be classified as a fast scrambler even though the exponential fits of the OTOCs are good fits. We see that this fitting procedure also results in the expected $a/\beta$ behavior expected from a fast scrambler confirming our expectations.



\section{Krylov Subspace Methods}
\label{appx:krylov}
Krylov methods are one of the widely used numerical techniques for dealing with large matrices, particularly obtaining their characteristic equations. However, for quantum mechanical systems, such methods have proven to be useful in studying time evolution of quantum states/operators without exactly diagonalizing the Hamiltonian for a system. The method is based on the Lanczos algorithm which can be summarized as follows~\cite{Sanchez-Garrido:2024pcy,lanczos1950iteration}, 
\begin{enumerate}
    \item Set $\ket{K_0} = \ket{v_0}$, where $\ket{v_0}$ is the initial state you start with.
    \item $a_0 = \langle K_0 \lvert H \rvert K_0\rangle$
    \item Loop over $n$, where $n \geq 1$
        \begin{enumerate}
            \item $\ket{A_n} = H\ket{K_{n-1}}$
            \item FRO : $\ket{A_n} \to \ket{A_n} - \sum_{m=0}^{n-1}\langle K_m \lvert A_n\rangle \ket{K_m}$
            \item $b_n = \sqrt{\langle A_n \lvert A_n\rangle}$. If $b_n = 0$, exit the loop else continue.
            \item $\ket{K_n} = \frac{1}{b_n}\ket{A_n}$
            \item $a_n = \langle K_n \lvert H \rvert K_n\rangle$
        \end{enumerate}
\end{enumerate}
This algorithm has numerical instabilities due to the round-off errors and it requires either a full or a partial re-orthogonalization to be carried out to ensure the orthogonality between Krylov vectors. This procedure is basically a  Gram-Schmidt process applied to Krylov vectors/operators after every iteration.  
Above algorithm can also be used for operators~\cite{Sanchez-Garrido:2024pcy} with redefining the inner products accordingly.  Once the Lancozs coefficients are obtained using this algorithm, we can write the recursion relations and solve them.

To obtain the time evolution of these vectors, recursion relations in Eq.~\ref{eq:rec_op} can be written in a vector-matrix linear ordinary differential equation form given in Eq.~\ref{eq:diffeq}. 
\begin{equation}
    \label{eq:diffeq}
    \frac{d\vec{\phi}}{dt} = M\vec{\phi}
\end{equation}
where $\vec{\phi} = (\phi_0, \phi_1, \ldots, \phi_{\mathbb{K}-1})^T$ and $M$ is a tridiagonal matrix which has a different form  for states $M_s$ and operators $M_o$. 

These matrices are used for obtaining state and operator complexities which are given by Eq.~\ref{eq:state_comp_mat} and Eq.~\ref{eq:op_comp_mat} respectively.
\begin{equation}
    \label{eq:state_comp_mat}
    M_s = -i\begin{bmatrix}
        a_0 & b_1 & 0 & 0 & \ldots \\
        b_1 & a_1 & b_2 & 0 &\ldots \\
        0 & b_2 & a_2 & b_3 & \ldots \\
        \vdots & \ddots & \ddots & \ddots & b_{\mathbb{K}-1}\\
        0 & 0 & \ldots & b_{\mathbb{K}-1} & a_{\mathbb{K}-1}
        
    \end{bmatrix}
\end{equation}

\begin{equation}
    \label{eq:op_comp_mat}
    M_o = \begin{bmatrix}
        0 & -b_1 & 0 & 0 & \ldots \\
        b_1 & 0 & -b_2 & 0 &\ldots \\
        0 & b_2 & 0 & -b_3 & \ldots \\
        \vdots & \ddots & \ddots & \ddots & -b_{\mathbb{K}-1}\\
        0 & 0 & \ldots & b_{\mathbb{K}-1} & 0
    \end{bmatrix}
\end{equation}

 The solution to the differential equation in Eq.~\ref{eq:diffeq} can be given as  $\vec{\phi}(t) = e^{Mt}\vec{\phi}(0)$ where $\vec{\phi}(0) = (1,0,0,\ldots,0)^T$ and is also expressed by the following recursion relation. 
\begin{equation}
    \dot{\phi}_n(t) = b_n\phi_{n-1}(t) -b_{n+1}\phi_{n+1}(t) 
\end{equation}
 
  A fourth-order Runge Kutta (RK4) method is then used to solve this recursion relation numerically. 

\end{appendix}

\section{Numerical data for Krylov Subspace Methods}
\label{appx:krylov_data}
In this section, we show (Table.~\ref{tab:opkrylov_fits}) the fit results from the exponential and linear fits to operator K-complexity and K-entropy. The complexity is fitted to an exponential function of the form $f(x) = a(e^{bx} - c)$ and the entropy is fitted to a linear function $f(x) = ax + b$.
\begin{table}[h!]
    \centering
    \begin{tabular}{|c|c|c|}
         \hline
         $\ell_{\rm max}$ & $b$ (complexity) & $a$ (entropy)\\
         \hline
         0.5 & 12.64(69) & 2.885(74)\\
         1.0 & 15.7(1.0) & 4.093(98)\\
         1.5 &  18.4(1.4) & 6.535(94)\\
         \hline
    \end{tabular}
    \caption{Fit parameter values for the argument of the exponential $b$ for the operator complexity and slope of the linear fit $a$ in operator entropy.}
    \label{tab:opkrylov_fits}
\end{table}

In the case of state K-complexity, Table.~\ref{tab:skrylov_fits} shows the value of the normalized peak in state complexity for different curvatures.
\begin{table}[h!]
    \centering
    \begin{tabular}{|c|c|}
         \hline
         $\ell_{\rm max}$ & $b$\\
         \hline
         0.5 & 1.1155\\
         1.0 & 1.1199\\
         1.5 & 1.1220\\
         \hline
    \end{tabular}
    \caption{Peak value of the normalized state complexity for different $\ell_{\rm max}$ values.}
    \label{tab:skrylov_fits}
\end{table}

\bibliographystyle{sn-mathphys-num}
\bibliography{hyp_ising}

\end{document}